\documentclass[prx,	twocolumn,showpacs,amsmath,amssymb,superscriptaddress,longbibliography]{revtex4-2}

\usepackage{color}
\definecolor{blues}{RGB}{0,108,169}
\definecolor{blaa}{RGB}{153,153,255}
\definecolor{blaa}{RGB}{0,0,125}
\definecolor{filtered}{RGB}{153,0,51}
\definecolor{raw}{RGB}{255,177,100}
\definecolor{filtered}{RGB}{175,0,0}
\definecolor{groen}{RGB}{0,150,0}
\definecolor{lil}{RGB}{160,33,160}
\definecolor{orang}{RGB}{255,60,0}
\usepackage{soul}

\usepackage{amssymb,amsmath}
\usepackage{siunitx}
\usepackage[colorlinks=true,urlcolor=BrickRed,citecolor=blues,linkcolor=ForestGreen]{hyperref}
\usepackage[resetlabels,labeled]{multibib}
\newcites{S}{References}

\newcommand{\wsixj}[6]{\ensuremath{\begin{Bmatrix} #1 & #2 & #3 \\ #4 & #5 & #6 \end{Bmatrix}}}

\usepackage{graphicx}
\usepackage{epstopdf}
\usepackage{float}
\usepackage[dvipsnames]{xcolor}
\newcommand{\tranz}{\!\!\leftrightarrow\!\!}
\newcommand{\tranzs}{\!\!\leftrightarrow\!\!}

\newcommand{\ket}[1]{|#1\rangle}
\newcommand\bra[1]{\left\langle#1\right|}
\newcommand{\braket}[3]{\ensuremath{\left<#1\right|#2\left|#3\right\rangle}}
\newcommand{\ReducedMat}[3]{\ensuremath{\left<\left. #1 \right\| #2 \left\| #3 \right.\right>}}

\newcommand{\eref}[1]{Eq.~(\ref{#1})}

\newcommand{\fref}[1]{Fig.~\ref{#1}}

\newcommand{\Fref}[1]{Figure~\ref{#1}}

\usepackage{orcidlink}
\usepackage{nicematrix}
\NiceMatrixOptions{
	code-for-first-row = \color{black!30!ForestGreen} ,
	code-for-last-row = \color{black!30!ForestGreen} ,
	code-for-first-col = \color{black!30!ForestGreen} ,
	code-for-last-col = \color{black!30!ForestGreen}
}

\makeatletter
\newcommand{\ORCID}[1]{{\orcidlink{#1}}}
\makeatother

\DeclareFontFamily{U}{mathb}{\hyphenchar\font45}
\DeclareFontShape{U}{mathb}{m}{n}{
	<5> <6> <7> <8> <9> <10> gen * mathb
	<10.95> mathb10 <12> <14.4> <17.28> <20.74> <24.88> mathb12
}{}
\DeclareSymbolFont{mathb}{U}{mathb}{m}{n}
\DeclareFontSubstitution{U}{mathb}{m}{n}
\DeclareFontFamily{U}{mathx}{\hyphenchar\font45}
\DeclareFontShape{U}{mathx}{m}{n}{
	<5> <6> <7> <8> <9> <10>
	<10.95> <12> <14.4> <17.28> <20.74> <24.88>
	mathx10
}{}
\DeclareSymbolFont{mathx}{U}{mathx}{m}{n}
\DeclareFontSubstitution{U}{mathx}{m}{n}
\DeclareMathDelimiter{\thickvert}{0}{mathb}{"7E}{mathx}{"1F}

\newcommand\ii{\text{i}}

\begin{document}
%\title{Rydberg atomic polarimetry of radio-frequency fields}
\title{EIT Spectroscopy of Rydberg Levels Dressed by Linearly Polarized RF fields: Complementary Angular Response for Two Types of Transition Ladders}
\author{Matthew Cloutman\ORCID{0009-0009-5746-8425}}
\affiliation{
	Department of Physics, QSO-Quantum Science Otago, and Dodd-Walls Centre for Photonic and Quantum Technologies,
	University of Otago, Dunedin 9016, New Zealand
}%
\author{Matthew Chilcott\ORCID{0000-0002-1664-6477}}
\affiliation{
	Department of Physics, QSO-Quantum Science Otago, and Dodd-Walls Centre for Photonic and Quantum Technologies,
	University of Otago, Dunedin 9016, New Zealand
}%
\author{Alexander Elliott}
\affiliation{
	Department of Physics and Dodd-Walls Centre for Photonic and Quantum Technologies,
	University of Auckland, New Zealand
}%
\author{J. Susanne Otto\ORCID{0000-0003-0760-3800}}
\affiliation{
	Department of Physics, QSO-Quantum Science Otago, and Dodd-Walls Centre for Photonic and Quantum Technologies,
	University of Otago, Dunedin 9016, New Zealand
}%
\author{Amita B. Deb\ORCID{0000-0002-2427-3500}}%
\affiliation{
	School of Physics and Astronomy, University of Birmingham, Edgbaston, Birmingham B15 2TT, United Kingdom
}%
\author{Niels Kj{\ae}rgaard\ORCID{0000-0002-7830-9468}}%
\email{niels.kjaergaard@otago.ac.nz}
\affiliation{
	Department of Physics, QSO-Quantum Science Otago, and Dodd-Walls Centre for Photonic and Quantum Technologies,
	University of Otago, Dunedin 9016, New Zealand
}%
\date{\today}

\begin{abstract}
  Rydberg atoms efficiently link photons between the radio-frequency (RF) and optical domains. They furnish a medium in which the presence of an RF-field imprints on the transmission of a probe laser beam by altering the coherent coupling between atomic quantum states. The immutable atomic energy structure underpins quantum-metrological RF-field measurements and has driven intensive efforts to realize inherently self-calibrated sensing devices. Here we investigate spectroscopic signatures owing to the quantization of atomic angular momentum. Using an electromagnetically-induced transparency (EIT) sensing scheme, specific combinations of atomic terms are shown to give rise to universal, distinctive fingerprints in the detected optical fields upon rotating a linearly polarized RF field. Employing a dressed state picture, we identify two types of atomic angular momentum ladders that display strikingly disparate spectroscopic characteristics, including the distinctive absence or presence of a central spectral EIT peak. Our study adds important insights into the prospects of Rydberg atomic gases for quantum metrological electric field characterization including polarimetry. In particular, it calls into question prevailing interpretations of SI-traceable Rydberg atom electrometers.

\end{abstract}

\maketitle
\section{Introduction}
Since the turn of the millennium, Rydberg atoms have experienced a veritable renaissance, emerging as a workhorse for quantum-technological applications \cite{Adams2019}. Prompted by the proposal to exploit their long-range dipolar interactions for quantum-computational gate operations \cite{Jaksch2000}, extensive efforts were devoted to making Rydberg atomic quantum information processors a reality~\cite{Bluvstein2023}. Along another avenue, Rydberg atoms found themselves at the heart of vapor-based sensing applications \cite{Anderson2020a,Fancher2021,Simons2021a,Yuan2023,Zhang2024} following the seminal 2012 work by Sedlacek et al. \cite{Sedlacek2012}.

Unlike Rydberg-based quantum computers and simulators, which require laser cooled atoms in an elaborate ultrahigh vacuum setup, Rydberg sensing can be achieved with simple, sealed glass cells. The cells contain a room-temperature gas of alkali atoms to form a transducer `wired up' with laser beams that read out measurements~\cite{Anderson2021,Otto2023}. These sensors capitalize on the spectacular magnitude of electric transition dipole moments between Rydberg excited states to accurately determine the amplitudes of RF electric fields with frequencies extending into the THz domain. Moreover, because such measurements can be tied to transitions between quantum states of an atom, Rydberg-atomic schemes establish an inherently self-calibrated RF-field probe~\cite{Holloway2014,Schmidt2024}.

 The obvious metrological advantage of self-calibrated, SI-traceable atomic probes over conventional, metallic dipole antennas has triggered a worldwide surge in experiments devoted to vapor-based RF-field sensing \cite{Jing2020,Otto2021a,Simons2021a,Cui2023,Liu2022,Schmidt2024}. The appealing idea of SI-traceability, where a spectroscopic measurement is linked to Planck's constant via an accurately calculated transition dipole moment, typically departs from the notion of an ideal atomic four-level ladder. In this scenario, a linking dipole moment can be unambiguously assigned. Real atoms, however, exhibit a multi-state structure including angular momentum states, and it therefore raises an important question as to how this should be accounted for.

Rydberg atomic vapor-cell sensing offers unique capabilities for measuring polarized RF-fields. For example, an isotropic response to a linearly-polarized field is fundamentally prohibited for a conventional electric receiver antenna \cite{Mathis1951,Mathis1954,Scott1966}. In contrast, by employing properly selected quantum states of a Rydberg atomic system, it may realise an ideal isotropic receiver, where a measurement of an incoming linearly polarized RF field is independent of its directions of propagation and polarisation \cite{Anderson2021,Cloutman2024,Chopinaud2024,Yuan:24}. Alternatively, quantum states can be selected to achieve a polarization-dependent EIT response that enables atom-based RF vector electrometry \cite{Sedlacek2013,Jiao2017,Wang:23,Elgee2024}. For example, Ref.~\citenum{Sedlacek2013} established an optical signal that would vary quasi-sinusoidally when rotating a microwave field.

In this study, we take a closer look at the mechanism giving rise to a polarization-dependent EIT signal for atoms with Rydberg levels dressed by a linearly polarized RF field.
Specifically, we identify two types of quantum-state ladders that display distinctive out-of-phase oscillatory optical read-out signals when the polarization angle of an incoming RF field is scanned.  Using a ``dressed state'' picture, we provide a framework for a physical understanding of the contrasting polarimetric responses we observe in both experiment and density matrix simulations. As a corollary, our work highlights that even in their simplest configuration with all fields linearly co-polarized, the system can generally not be reduced to an effective four-level ladder of atomic levels. This more or less tacit assumption underpins the majority of reported realizations of self-calibrated electrometers which aim to link electric field measurements to a single, well-defined atomic dipole moment.

\vspace{0.3cm}

\section{Theory}
\subsection{Polarization dependent RF sensing in Rydberg EIT---a conceptual model}
\begin{figure}[t!]
	\centering
	\includegraphics{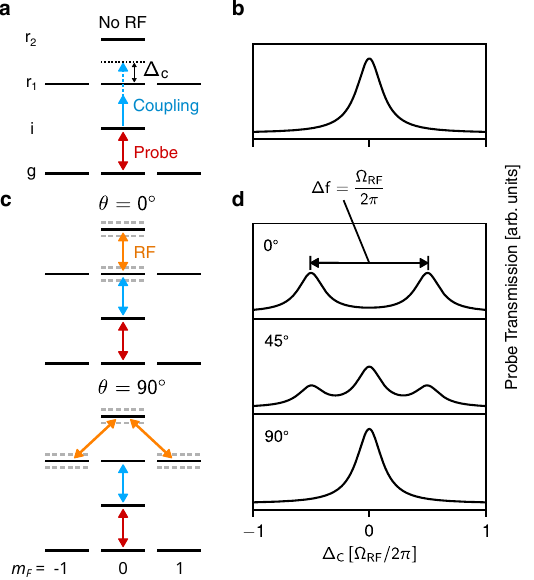}
	\caption{(a) Level diagram of a four-level model atom with three-fold degeneracy of its ground ($g$) and second excited level ($r_1$). A linearly-polarised probe field is resonant with the transition from $g$ to the first excited level $i$. A coupling field has a frequency close to the transition from $i$ to $r_1$, giving rise to an EIT peak for the transmitted probe field when its detuning $\Delta_\text{c}$ is scanned (b). (c) The model atom dressed in an RF-field that is parallelly ($\theta=0^\circ$) or perpendicularly ($\theta=90^\circ$) polarized with respect to the optical probe and coupling fields. The resulting AT splitting is shown indicated with dashed lines. {(d)} Simulated probe transmission spectra for $\theta=0^\circ,45^\circ,90^\circ$. $\Omega_\text{RF}$ denotes the Rabi frequency characterizing the RF coupling.}
	\label{fig:mickey-mouse}
\end{figure}
To elucidate how an EIT signal might depend on the polarization of an applied RF field, we introduce in \fref{fig:mickey-mouse}(a) the energy level diagram for a fictitious four-level model atom that includes angular momentum degeneracy for the ground and second excited levels. A probe laser field linearly polarized along the $z$-axis matches the optical transition frequency between the ground level $g$ and an excited level $i$; choosing $\bf\hat{z}$ as the quantization axis for the model atom, the probe field can drive a $\pi$-transition between $\ket{g,m=0}$ (one of three degenerate angular momentum states of the ground level) and $\ket{i,m=0}$.

An optical probe beam interacting resonantly with a collection of model atoms on the $g\tranzs i$-transition will be attenuated as light is scattered out of the probe mode. However, by adding an additional laser field that couples the excited level $i$ resonantly to a third level $r_1$, the atomic medium can be made transparent to the probe laser beam through the mechanism of electromagnetically induced transparency (EIT) \cite{Fleischhauer2005}. For our model atom, the excited intermediate state $\ket{i,m=0}$ connects with a higher-lying state $\ket{r_1,m=0}$ (one of three degenerate angular momentum states of the Rydberg level $r_1$) through an optical coupling field that is linearly co-polarized with the probe field along $\bf\hat{z}$. Because of the ascending energies of the levels involved, the resulting EIT for the probe field is referred to as a ladder scheme. The EIT transmission is maximized for a coupling field detuning $\Delta_\text{c}=0$ as illustrated in \fref{fig:mickey-mouse}(b).

We next consider the effect of adding a linearly-polarized RF field to our model atom that resonantly couples the Rydberg level $r_1$ to another Rydberg level $r_2$. It is unimportant if $r_2$ lies above or below $r_1$ in energy and without loss of generality we will depict it above $r_1$ [see \fref{fig:mickey-mouse}(a)]. For the RF field polarization aligned at an angle $\theta=0^\circ$ with respect to the optical fields, the non-degenerate state $\ket{r_2,m=0}$ connects to $\ket{r_1,m=0}$, the top rung of the EIT ladder. The RF coupling $\ket{r_2,m=0}\tranzs\ket{r_1,m=0}$ acts to disrupt the optical EIT for $\Delta_\text{c}=0$ through the Autler-Townes (AT) effect \cite{CohenTannoudji1996}. For $\theta=90^\circ$, the $\ket{r_1,m=0}$ top rung of the EIT ladder is left unconnected by the RF-field and EIT proceeds exactly as if the RF field had not been present. \Fref{fig:mickey-mouse}(c) shows level diagrams and the coupled states for the two cases $\theta=0^\circ,90^\circ$; the state-splitting resulting from the coupling is indicated with dashed lines. \Fref{fig:mickey-mouse}(d) displays the associated EIT spectra of the optical probe transmission versus the coupling laser field detuning $\Delta_\text{c}$. This is double-peaked (AT doublet) for $\theta=0^\circ$ and single-peaked for $\theta=90^\circ$ while a triple-peaked spectrum is found at $\theta=45^\circ$ (all spectra for angles between $\theta=0^\circ$ and $\theta=90^\circ$ will have three peaks, with the height ratio between the central peak and a side lobe increasing with $\theta$).

\begin{figure*}
 	\centering
 	\includegraphics[width=\textwidth]{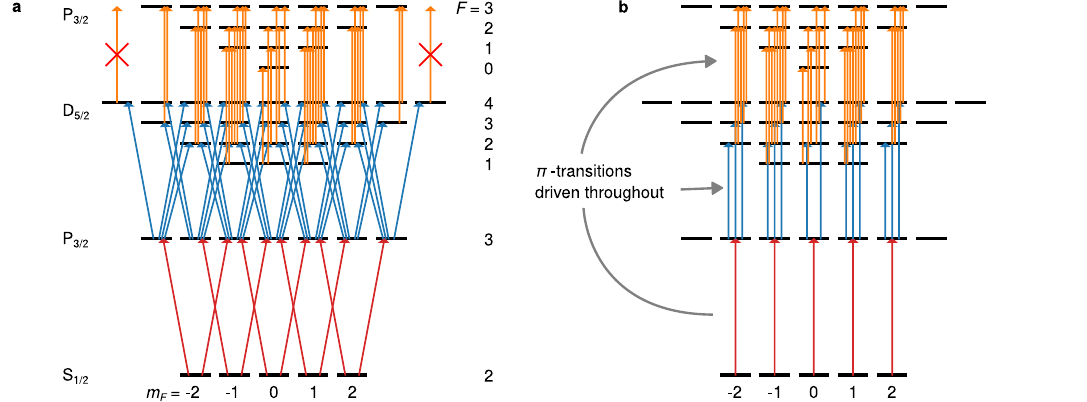}
 	\caption{Hyperfine level diagrams for the $S_{1/2}^{F=2}\tranz P_{3/2}^{F=3}\tranz D_{5/2}\tranz P_{3/2}$ excitation ladder with the ultimate and penultimate levels (Rydberg levels) resonantly coupled by an RF field, which is linearly polarized along the atomic quantization axis and drives $\pi$-transitions (orange arrows). Red and blue arrows show the allowed transitions for optical probe and coupling fields linear polarized (a) perpendicularly to the RF field (optical $\sigma$-transitions) and (b) parallel to the RF field (optical $\pi$-transitions). The $F$-levels for the Rydberg states are energetically degenerate, and have been offset vertically to illustrate the dipole-allowed transitions in play.}
 	\label{fig:hyperfine-wide}
 \end{figure*}
 
 \vspace{5mm}
\subsection{Rydberg atomic RF-polarimetry in a complex multi-level atom}
The simple EIT-probed model-atom level structure of \fref{fig:mickey-mouse}(a) displays two salient features when a linearly-polarized RF field is added. For the polarization $\theta=0^\circ$, an AT doublet is encountered in the spectrum due to a vertical, pure $m=0$ four-level ladder, while for $\theta=90^\circ$ we have a pure three-level system and a single central EIT peak. As such, the EIT spectrum carries an imprint of the polarization state of the RF field. A similar polarization dependence will generally be encountered for real atoms with a more complex level structure and this opens up the prospect of atomic RF polarimetry.  Its utilization, however, necessitates a detailed understanding of the polarization signature resulting from a particular atomic level structure. 

\Fref{fig:hyperfine-wide}a reproduces the essence of Fig. 1(a) of Ref.~\citenum{Sedlacek2013} that considered the $5S_{1/2}^{F=2}\tranz 5P_{3/2}^{F=3}\tranz53D_{5/2}\tranz54P_{3/2}$ ladder of $\rm^{87}Rb$ as a tool for vector electrometry. The hyperfine sub-levels ($F$-levels) of the $53D_{5/2}$ level and the upper $54P_{3/2}$ are unresolved, so to enable a visual tracking of all dipole-allowed transitions in the diagram, relative vertical offsets ($\propto F$) have been added to the Rydberg sub-states \cite{niels}. The red crosses serve to illustrate that an RF field polarized along the chosen quantization axis does not directly connect the stretched states
$D_{5/2}(F=4,m_F=\pm4)$ to the upper $\rm P_{3/2}$ Rydberg level. Reference \citenum{Sedlacek2013} points out that the system when probed with optical fields polarized perpendicularly to the RF field---the situation encountered in \fref{fig:hyperfine-wide}(a)---therefore presents three-level EIT excitation pathways [specifically, $S(F=2,m_F=\pm2)\rightarrow P(F=3,m_F=\pm 3)\rightarrow D(F=4,m_F=\pm 4)$]. Reference \citenum{Schlossberger2024} reiterates this point, noting that the $D_{5/2}(F=4,m_F=\pm 4)$ states are ``unsplit'' by the RF field. The EIT spectrum for crossed RF and optical polarizations ($\theta=90^\circ$) will hence include a central peak at zero detuning as either the probe or coupling field is scanned.
 \begin{figure}[b!]
    \centering
    \includegraphics{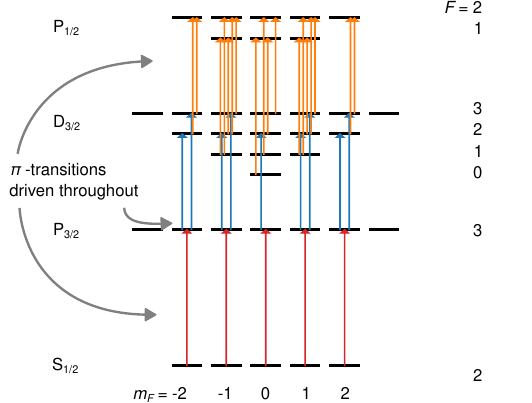}
    \caption{Hyperfine level diagrams for the $S_{1/2}^{F=2}\tranz P_{3/2}^{F=3}\tranz D_{3/2}\tranz P_{1/2}$ excitation ladder with the ultimate and penultimate levels (Rydberg levels) resonantly coupled (orange arrows) by an RF field, which is linearly polarized along the atomic quantization axis. The optical coupling (blue arrows) and probe (red arrows) fields are also linearly polarized along the quantization axis and $\pi$-transitions are driven throughout the system.}
    \label{fig:pitrans2}
\end{figure}
 \begin{figure*}[t!]
 	\centering
 \includegraphics[width=\textwidth]{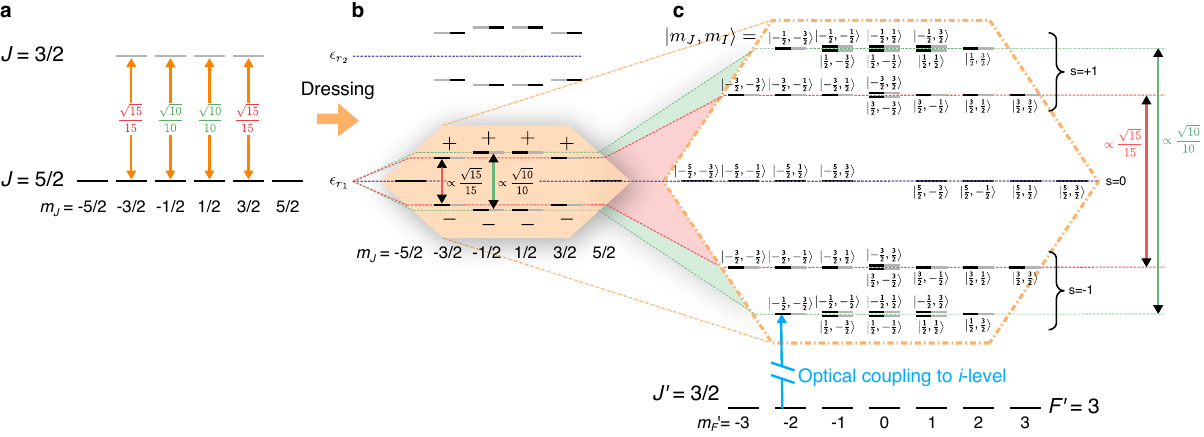}
 	\caption{RF-dressed Rydberg levels---procedure for finding the energy structure probed via EIT. (a) A linearly-polarized RF field couples states of lower ($J=5/2$, black) and upper ($J=3/2$, grey) Rydberg levels via $\pi$-transitions with dipole matrix elements as indicated. (b) Field-dressed Rydberg levels with $m_J$-dependent AT-splitting. Each grey-black state in the diagram represents an equal two-component superposition of $J=5/2$ and $J=3/2$ states with the adjacent $\pm$ indicating its symmetry. (c) Labelling the lower dressed manifold of (b) to include all possible values for the nuclear spin projection $m_I$ so that each state 
    inside the hexagon is designated by $\ket{m_J,m_I}$ (and hence $m_F=m_J+m_I$) as well as a symmetry parameter $s$. Each $\ket{m_J,m_I;s}$ can be expanded using \eref{eq:expansion}, which facilitates the evaluation of the transition strength for the optical coupling from a state 
    $\ket{J'=3/2,F'=3,m_F'=m_F}$ through \eref{eq:3j}.}
 	\label{fig:r1r2} 
 \end{figure*}
 
 The scenario of parallel RF and optical polarizations is captured in \fref{fig:hyperfine-wide}(b) \cite{mini}. Here, Ref.~\citenum{Sedlacek2013} argues that ``In this case, $\pi$ transitions are driven throughout the system and all excitation pathways experience a four-level system. The theoretical and experimental spectra have two transmission peaks...''. Our theoretical treatment will, as we shall see, in fact predict four transmission peaks for a $S_{1/2}^{F=2}\tranz P_{3/2}^{F=3}\tranz D_{5/2}\tranz P_{3/2}$ ladder system; the reason that only two peaks are observed in Ref.~\citenum{Sedlacek2013} is because insufficient RF power is applied to resolve the structure in the presence of line broadening mechanisms. But more importantly, and at the heart of understanding the physics in play, the absence of a central transmission peak is \textit{not} the result of $\pi$-transitions being driven throughout the system! As a case in point, we will contrast the $S_{1/2}^{F=2}\tranz P_{3/2}^{F=3}\tranz D_{5/2}\tranz P_{3/2}$ system (which we shall call type-I) to that of $S_{1/2}^{F=2}\tranz P_{3/2}^{F=3}\tranz D_{3/2}\tranz P_{1/2}$ (type-II). \Fref{fig:pitrans2} shows a level diagram of the latter with transitions for co-polarized RF and optical fields. Similarly to the $S_{1/2}^{F=2}\tranz P_{3/2}^{F=3}\tranz D_{5/2}\tranz P_{3/2}$ system in \fref{fig:hyperfine-wide}b, only $\pi$-transitions are driven throughout the system, yet, as we shall demonstrate, the $S_{1/2}^{F=2}\tranz P_{3/2}^{F=3}\tranz D_{3/2}\tranz P_{1/2}$ system displays a dominant central transmission peak for $\theta=0^\circ$. In fact, the central transmission peak encountered at $\theta=0^\circ$ dwarfs the central peak at $\theta=90^\circ$.
 
 \vspace{3mm}
\subsection{Criterion for a central EIT peak}
\subsubsection{Type-I system: absent central EIT peak for co-polarized fields}
\label{sec:type1}
\noindent To seek the actual underlying origin of the absent central EIT peak of the system of \fref{fig:hyperfine-wide}(b)
\begin{equation}
	\underbrace{S_{1/2}^{F=2}}_{g}\tranz\underbrace{P_{3/2}^{F=3}}_{i}\tranz\underbrace{D_{5/2}}_{r_1}\tranz\underbrace{P_{3/2}}_{r_2}, \tag{I}\label{eq:typeI}
\end{equation}
we first consider, in isolation, the two Rydberg levels $r_1$ and $r_2$ at energies $\epsilon_{r_1}$ and $\epsilon_{r_2}$, respectively, dressed by a resonant linearly-polarized RF field with angular frequency $\omega_{\rm RF}=|\epsilon_{r_2}-\epsilon_{r_1}|/\hbar$. As mentioned above, hyperfine structure is completely negligible for the Ryd\-berg levels, and this part of the problem can therefore be treated in a fine-structure basis where it separates into a series of four individual, driven two-level systems and two uncoupled $r_1$ spectator states ($m_J=\pm5/2$). The Rabi frequency for each two-level system depends on $|m_J|$ and varies as indicated in \fref{fig:r1r2}(a). As a result of the resonant RF coupling, the AT effect leads to varying splittings across the lower $r_1$-level as shown in \fref{fig:r1r2}(b). The energy spectrum for the $\epsilon_{r_1}$-level now has five peaks where, for the bare atom, there was only one---this results directly from the  $\updownarrow_{J_1=5/2}^{J_2=3/2}$-level structure \footnote{A general rule for the number of spectral peaks is given in Ref.~\citenum{Cloutman2024}}.

\begin{figure*}
    \centering
    \includegraphics{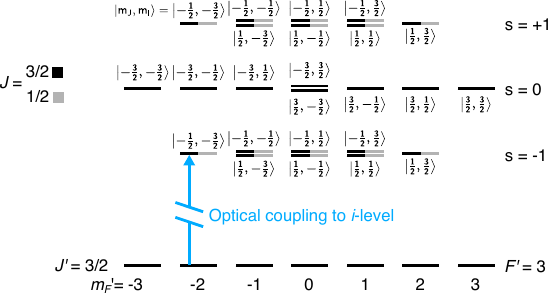}
    \caption{Diagram for evaluating the optical coupling from a state $\ket{J'=3/2,F'=3,m_F'=m_F}$ to a state in a field-dressed $D_{3/2}\tranzs P_{1/2}$ Rydberg manifold. The procedure for obtaining this diagram follows that of \fref{fig:r1r2} and each of the states $\ket{m_J,m_I;s}$ can be expanded using \eref{eq:expansion}.}
    \label{fig:leveldiag4}
\end{figure*}

 Unlike the Rydberg levels $r_1$ and $r_2$, the $g$ and $i$ levels display resolvable hyperfine structure, and excitation to $r_1$ specifically departs from the $F=3$ $i$-level. It is important to note that because of Laporte's rule, an $i$-level state cannot couple to the $r_2$-component of a dressed state because the change in orbital angular momentum between  $i$ and $r_2$ will then differ from unity (for our experiments both are described by spectroscopic $P$-terms). Hence we will exclusively be concerned with the transitions from states $\ket{F=3,m_F=-3}...\ket{F=3,m_F=3}$ of the $i$-level to the $r_1$-components of the dressed manifold centred on $\epsilon_{r_1}$.
 
 To establish how the RF field-dressing of the atom affects the optical EIT probing, we transform to a hyperfine basis, where $J$ is coupled to the spin of the atomic nucleus $I=3/2$ to form $F=J+I$. \Fref{fig:r1r2}(c) recasts the lower dressed manifold of \fref{fig:r1r2}(b) (orange hexagon) to include the possible projections $m_I$ of the nuclear spin. In doing so, each state of \fref{fig:r1r2}(c) is designated by quantum numbers $m_J$ and $m_I$ in addition to a parameter $s=0,\pm1$ that tracks the symmetry of the parental dressed state of \Fref{fig:r1r2}(b) from which it descends: every dressed state in \fref{fig:r1r2}(b) is either an equal symmetric ($s=+1$) or antisymmetric ($s=-1$) superposition of $r_1$ and $r_2$ components. The $r_1$ spectator states that are unaffected by the dressing field ($m_J=\pm5/2$) are labeled by $s=0$.
Expanding the $r_1$-part (i.e., $J=5/2$) of each state of \fref{fig:r1r2}(c) as
\begin{equation}\label{eq:expansion}
	\ket{m_Jm_I;s}=\left[\frac{1}{\sqrt{2}}\right]^{|s|}\sum_{F=1}^4\underbrace{\bra{JIFm_F}Jm_JIm_I\rangle}_{C^{Fm_F}_{Jm_JIm_I}} \ket{Fm_F},
\end{equation}
we obtain the required transformation to the hyperfine basis.

Using the expansion \eref{eq:expansion}, a measure for the strength of an optical $\pi$-transition from a particular state $\ket{J'F'm_F'}$ of the $i$-level (i.e., $J'=3/2$) to a state of the lower dressed Ryd\-berg manifold can now be found (see Supplementary Note 1 \cite{suppcite}):
\begin{eqnarray}\label{eq:3j}
&\phantom{.}&|\braket{J'F'm_F'}{-er_0}{Jm_JIm_I;s}|^2\nonumber\\&\phantom{.}&= \frac{1}{2^{|s|}}e^2\left|\sum_{F=1}^4 C^{Fm_F}_{Jm_JIm_I} \braket{F'm_F'}{r_0}{Fm_F}\right|^2\nonumber\\&\phantom{.}& \propto 2^{-|s|}\!\begin{pmatrix}
    J' & J & 1\\
    -m_J & m_J & 0
    \end{pmatrix}^2\!\!
    \begin{pmatrix}
    J' & F' & I\\
    -m_J & m_F' & m_J-m_F'
    \end{pmatrix}^2\!\!.
\end{eqnarray}

The first 3-$j$ symbol of \eref{eq:3j} vanishes if $|m_J|>J'$. An optical field co-polarized with the linear RF dressing field can therefore not couple any states of the $P_{J'=3/2}$ $i$-level manifold to the dressed Rydberg states with $m_J=\pm 5/2$---the central states in \fref{fig:r1r2}(c). As a result, the EIT spectrum obtained from probing the RF-dressed Rydberg states will not display a central peak.
\subsubsection{Type-II system: dominant central EIT peak for co-polarized fields}
Using the $S_{1/2}^{F=2}\tranz P_{3/2}^{F=3}\tranz D_{5/2}\tranz P_{3/2}$ system (type-\ref{eq:typeI} ladder) as an illustration, we have shown that, generally, for co-polarized fields the EIT spectrum will not have a central peak if $J_{r_1}>J_i$. Furthermore, an obvious requirement for a central peak is $J_{r_1}>J_{r_2}$---the $r_1$-level needs to have spectator states. An equivalent analysis of the superficially similar system
\begin{equation}
	\underbrace{S_{1/2}^{F=2}}_{g}\tranz\underbrace{P_{3/2}^{F=3}}_{i}\tranz\underbrace{D_{3/2}}_{r_1}\tranz\underbrace{P_{1/2}}_{r_2},\tag{II}\label{eq:typeII}
\end{equation}
presented in \fref{fig:pitrans2} yields the dressed level diagram shown in \fref{fig:leveldiag4}. When, for this type-\ref{eq:typeII} ladder, we consider the dipole matrix elements for optical $\pi$-transitions from the $i$-level to the Rydberg manifold, we see that $|m_J|=\{\tfrac{1}{2},\tfrac{3}{2}\}\leq J'=\tfrac{3}{2}$. We therefore do not experience the vanishing of the first $3$-$j$ symbol in \eref{eq:3j}. Rather, its common, non-zero squared magnitude 
\begin{equation}
	\begin{pmatrix}
    J' & J & 1\\
    -m_J & m_J & 0
    \end{pmatrix}^2\propto m_J^2
\end{equation}
for the central $m_J=\pm3/2$-states is ninefold that of the split $m_J=\pm1/2$ states. Hence, despite the fact that ``$\pi$ transitions are driven throughout the system'', a double-peak structure is \textit{not} predicted, but instead a triple-peak spectrum with a strong central peak.
\section{Experiment}
 \begin{figure}[t!]
    \centering
    \includegraphics{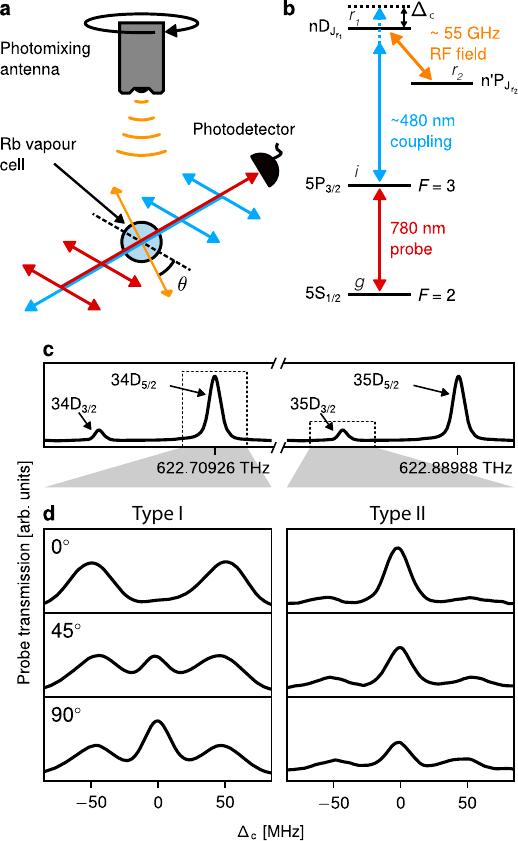}
    \caption{(a) Schematic of experimental setup. Probe and coupling lasers are linearly polarized in the horizontal plane and counter-propagate through a spherical Rb vapor cell. A photomixing antenna mounted in a motorized rotation stage emits linearly-polarized RF radiation along a vertical direction. (b) Atomic level structure for $\rm^{87}Rb$ applicable to both type-I and type-II ladders. (c) Probe transmission spectra showing EIT peaks when the coupling laser frequency is scanned. (d) Probe transmission spectra when resonant RF-radiation is applied to the $34D_{5/2}\tranzs35P_{3/2}$ (left column) and $35D_{3/2}\tranzs 36P_{1/2}$ (right column) transitions as a function of the coupling laser detuning $\Delta_\text{c}$; spectra for $\theta=0^\circ,\theta=45^\circ,90^\circ$ are shown.}
    \label{fig:setup}
\end{figure}
\noindent To verify the predicted complementarity in the polarization response of the two types of ladders encountered above,
we perform optical EIT spectroscopy on the $5S_{1/2}^{F=2}\tranz5P_{3/2}^{F=3}\tranz34D_{5/2}\tranz35P_{3/2}$ (type-\ref{eq:typeI}) and $5S_{1/2}^{F=2}\tranz5P_{3/2}^{F=3}\tranz35D_{3/2}\tranz36P_{1/2}$ (type-\ref{eq:typeII}) ladders of $\rm ^{87}Rb$.

\Fref{fig:setup}(a) shows a schematic of our experimental setup. Using a photodetector, we measure the transmission of a $\sim$780~nm probe laser beam, propagating along a horizontal axis through a centimeter-sized borosilicate glass cell containing a room-temperature rubidium vapor. The frequency of the probe laser is fixed and resonant with the $5S_{1/2}^{F=2}\tranzs5P_{3/2}^{F=3}$ D2 line of $\rm ^{87}Rb$ [see \fref{fig:setup}(b)]. A counter-propagating  $\sim$480~nm coupling laser is scanning in frequency across the transitions from  $5P_{3/2}\ (F=3)$ to either $34D_{5/2}$ (type-\ref{eq:typeI} ladder) or $35D_{3/2}$ (type-\ref{eq:typeII} ladder). 

\Fref{fig:setup}(c) shows the transmitted probe light for a frequency scan of the coupling light over the $i\leftrightarrow r_1$ transitions we utilize in our EIT sensing scheme (see appendix \ref{app1}). Both probe and coupling beams are linearly polarized in the horizontal plane. A photomixing antenna (see appendix \ref{app2}) emits linearly-polarized $\sim 55$~GHz RF-radiation down onto the vapor cell along a vertical axis intersecting the horizontally propagating optical beams. At the point of intersection, the angle between polarizations of the RF field and the optical fields is $\theta$ [see \Fref{fig:setup}(a)]. We can vary this angle by rotating the photoconductive antenna about its axis.

\begin{figure*}
    \centering
    \includegraphics[width=\linewidth]{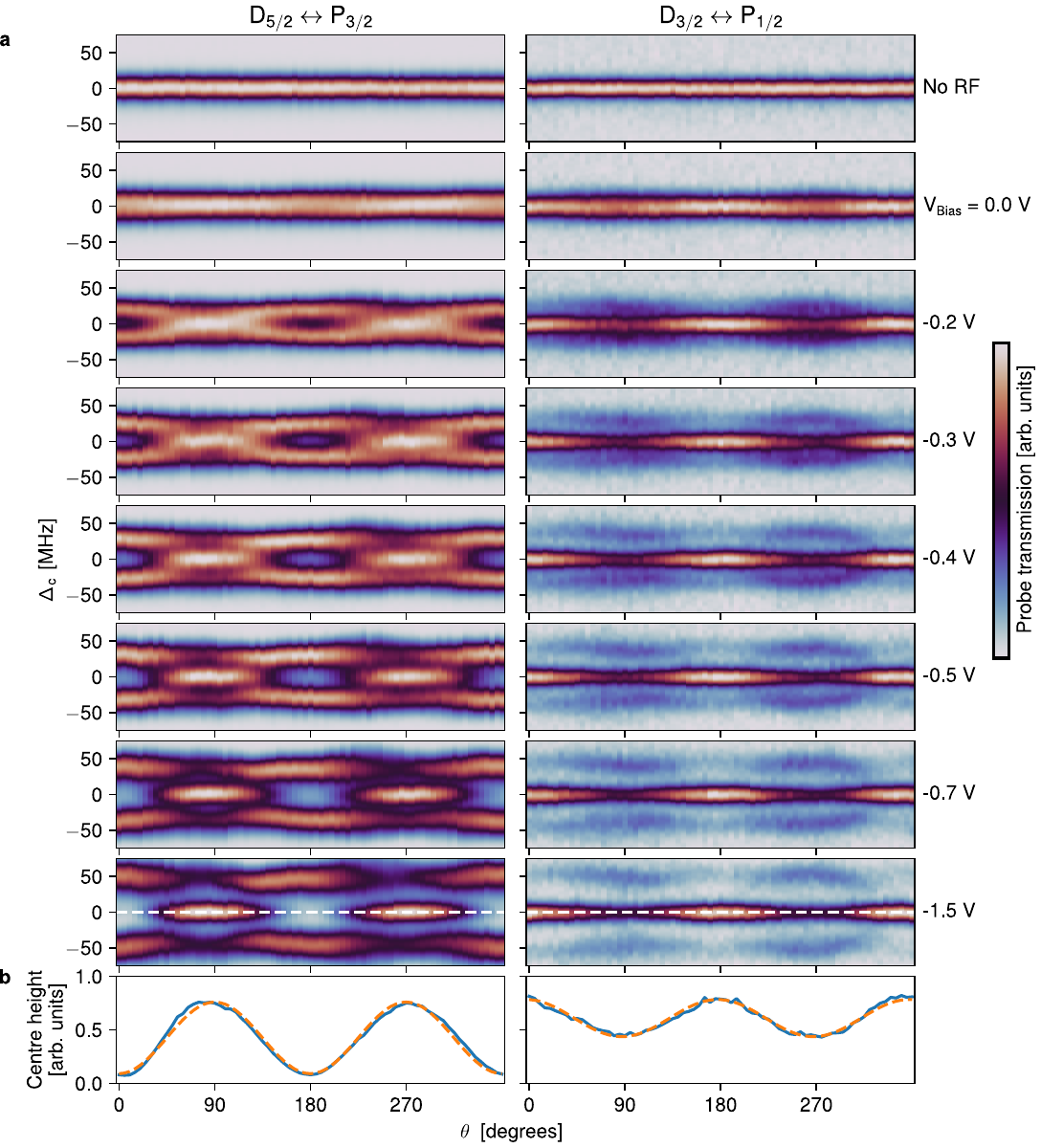}
    \caption{(a) Spectrograms from EIT-probing the RF-dressed $34D_{5/2}\tranzs 35P_{3/2}$ (left column) and $35D_{3/2}\tranzs 36P_{1/2}$ (right column)  transitions. The transmission of the probe field is measured on a parameter space spanned by the RF polarization angle $\theta$ and the coupling laser detuning $\Delta_\text{c}$ (see \fref{fig:setup}). From the top row, which is a reference spectrogram with no applied RF-field, moving down the RF power increases; the RF power shows a monotononically increasing, albeit nonlinear dependence with the bias voltage $V_\text{Bias}$ of the photomixing antenna. (b) Transmitted probe light on the $\Delta_\text{c}=0$ axis of the $V_\text{Bias}=-1.5$~V spectrograms as a function of $\theta$ (blue lines); sinusoidal fits are shown as dashed orange lines.}
    \label{fig:full-scans-transmission}
\end{figure*}

Figure \ref{fig:setup}(d) shows EIT spectra acquired for the resonantly driven $34D_{5/2}\tranzs 35P_{3/2}$ (left panel, type-\ref{eq:typeI} ladder) and $35D_{3/2}\tranzs 36P_{1/2}$ (right panel, type-\ref{eq:typeII} ladder) Rydberg transitions, for $\theta = 0^\circ$, $45^\circ$, and $90^\circ$. The polarization signatures for the type-\ref{eq:typeI} ladder in the left panel are similar to those reported for the $53D_{5/2}\tranzs 54P_{3/2}$ Rydberg transition in Ref.~\citenum{Sedlacek2013}. In particular, for $\theta = 0^\circ$ we note the absence of a central peak and what appears to be an AT doublet. Our dressed-picture analysis of the type-\ref{eq:typeI} system in Sec.~\ref{sec:type1} predicted an effective four-peak spectrum (a five-peak spectrum with no central component) at $\theta = 0^\circ$. However, the differential splitting of the Doppler-broadened $m_J=\pm 3/2$ and $m_J=\pm 5/2$ lines is too small at our applied RF power to be resolved. Rotating the emitter to $\theta=90^\circ$, we observe, again in correspondence with Ref.~\citenum{Sedlacek2013}, how the heights of the split peaks reduce while a prominent central peak emerges at $\Delta_\text{c} = 0$. Also for this situation, each of the split peaks in fact contains two spectral features that are obscured in the experiments by broadening.
\begin{figure*}
    \centering
    \includegraphics[width=1.5\columnwidth]{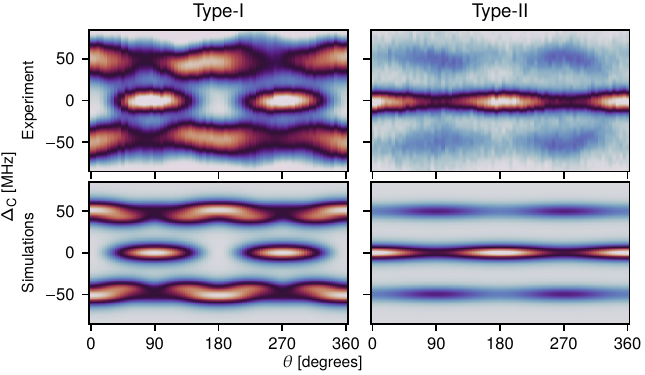}
    \caption{Simulated spectrograms (bottom row) for type-\ref{eq:typeI} (left column) and type-\ref{eq:typeII} ladders (right column). The corresponding experimental measurements ($V_\text{Bias}=-1.5$~V) are shown for comparison (top row).}
    \label{fig:simvsexp}
\end{figure*}

Comparing the EIT spectra of the RF-dressed type-\ref{eq:typeI} ladder to that of the type-\ref{eq:typeII} ladder, we note a compelling difference. In particular, for the type-\ref{eq:typeII} case and $\theta = 0^\circ$, where ``$\pi$-transitions are driven throughout the system'', we observe a prominent central peak at $\Delta_\text{c} = 0$ (see \fref{fig:setup}d). Moreover, in contrast to the $5S_{1/2}^{F=2}\tranz5P_{3/2}^{F=3}\tranz34D_{5/2}\tranz35P_{3/2}$ ladder, the central peak is minimized at $\theta = 90^\circ$.

The complementarity in spectral behaviour for type-\ref{eq:typeI} and type-\ref{eq:typeII} ladders is elucidated further in the `spectrograms' of \fref{fig:full-scans-transmission}, representing the sequentially recorded EIT spectra when rotating the RF-emitter in $5^\circ$ increments for a full $360^\circ$ revolution. The left and the right columns compare the type-\ref{eq:typeI} and -\ref{eq:typeII} cases as the power of the resonant RF dressing field is varied through the applied bias voltage to the photomixing antenna $V_\text{Bias}$. In both cases we observe an evolution where the $\Delta_\text{c}=0$ EIT peak (horizontal line features in the top row) progressively `breaks up' as an RF field is introduced and its power increased. The break-up happens in a complementary fashion, with the type-\ref{eq:typeI} ladder developing transmission minima and maxima at the $(\theta, \Delta_\text{c})$-coordinates where the type-\ref{eq:typeII} ladder presents its respective maxima and minima. This happens along the line $\Delta_\text{c}=0$, and \fref{fig:full-scans-transmission}b shows the corresponding out-of-phase $\pi$-periodic undulations.	
\section{Numerical simulations}
 The transmitted probe laser intensity $I$ in the spectrograms of \fref{fig:full-scans-transmission} is described by Beer-Lambert's law
 
 \begin{equation}\label{eq:beer}
    I(\theta,\Delta_\text{c}) = I_0 e^{-\alpha(\theta,\Delta_\text{c}) \ell},
\end{equation}
 where $I_0$ is the intensity of the probe light before traversing the distance $\ell$ through the vapor. The extinction coefficient $\alpha$ is proportional to a weighted sum over the off-diagonal elements of the density matrix that involves the combination of $g$ and $i$ -states [see \eref{eq:alpha} of Supplementary Note 2]. To simulate the variation of $\alpha(\theta,\Delta_\text{c})$ over a spectrogram for a given RF dressing power, we therefore calculate the steady-state atomic density matrix under the combined influence of the optical and RF fields (see appendix \ref{app3}). The density matrix for each case includes all individual $\ket{Fm_F}$ hyperfine states of the ladder levels (see \fref{fig:hyperfine-wide} and \fref{fig:pitrans2}). For the type-\ref{eq:typeI} ladder of \fref{fig:hyperfine-wide} involving a $\mathrm{D}_{5/2}\tranzs \mathrm{P}_{3/2}$ Rydberg transition, this amounts to 52 atomic states, whereas the type-II ladder of \fref{fig:pitrans2}, capped by the $35\mathrm{D}_{3/2}\tranzs 36\mathrm{P}_{1/2}$ transition, requires 36 atomic states. As in Ref.~\citenum{SedlacekThesis} an auxiliary ``dummy state'' is included to incorporate effective incoherent decay from transit-time broadening, and radiative decay from the Rydberg states to states not explicitly included in the model.
 
 In our simulations, the RF power is the only fitted quantity, and other input parameters to the modelling are either directly measured, as in the case of optical beam waists and powers, or they are tabulated values, as in the case of $^{87}$Rb atomic properties.
 \Fref{fig:simvsexp} showcases our ability to reliably simulate spectrograms for both type-\ref{eq:typeI}  and type-\ref{eq:typeII}  ladders.
\section{Discussion}
\subsection{Atomic standards: ideal four-level ladder versus real multilevel atoms}
\noindent The immense success of atomic clocks relies on the fact that the transition frequency between unperturbed quantized atomic states remains completely identical from atom to atom \cite{Lombardi2007}. This ensures quantum-enabled accuracy for cesium clocks at either end of the world ticking at exactly the same rate. In a similar quantum-metrological vein, Rydberg electrometry also aims at capitalising on atomic immutability and dynamics governed by quantum mechanics. As highlighted by Ref.~\citenum{Holloway2014}, it can quantify an electric field in SI-units from a frequency measurement, fundamental constants, and fixed, known atomic properties.

\begin{figure*}[t!]
    \centering
\includegraphics[width=1.5\columnwidth]{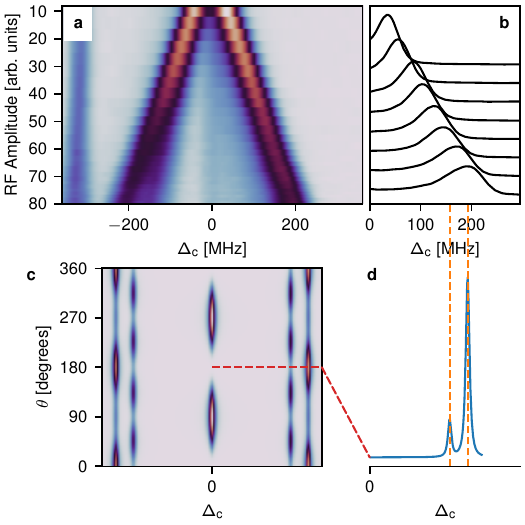}
    \caption{Electrometry using a type-\ref{eq:typeI} system with co-polarized fields. (a) Experimentally observed AT splitting of the $34D_{5/2}$ level centred on $\Delta_\text{c}=0$. The feature at $\Delta_\text{c}\sim -306$~MHz is the $34D_{3/2}$ level [see \fref{fig:setup}(c)]. Interaction with this state influences the $\Delta_\text{c}<0$ AT peak. (b) Development in the $\Delta_\text{c}>0$ AT peak with RF power increasing (top to bottom) from 10 to 80 (arb. units) in uniform steps. (c) Simulated spectrogram of the probe field transmission with artificially reduced broadening. (d) Simulated probe field transmission for co-polarized fields ($\theta=180^\circ$) with artificially reduced broadening.}
    \label{fig:fourfive}
\end{figure*}

\begin{figure}
    \centering
\includegraphics[width=\columnwidth]{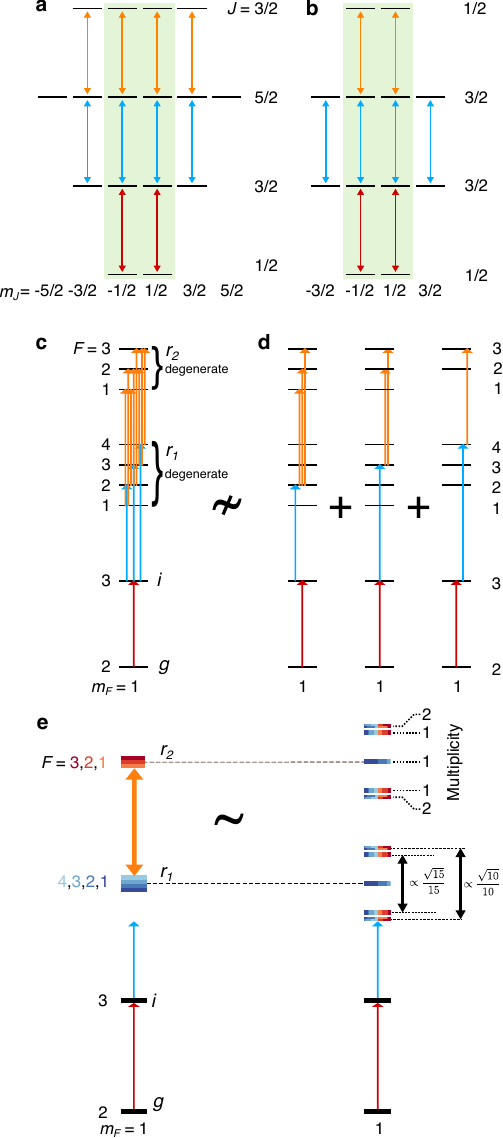}
    \caption{Level diagrams for a $J=\frac{1}{2}\frac{3}{2}\frac{5}{2}\frac{3}{2}$ system (a) and a $J=\frac{1}{2}\frac{3}{2}\frac{3}{2}\frac{1}{2}$ system (b) driven by $\pi$-transitions. Both realise two parallel ideal four-level ladders (green background). (c) $m_F=1$ ladder of \fref{fig:hyperfine-wide}(b) in isolation. The system cannot be broken down into parts as shown in (d), i.e., (c)$\nsim$(d). (e) Level splittings obtained from diagonalizing the $7\times 7$ matrix describing the RF-coupling (orange arrow) amongst the $F$-states of the Rydberg levels $r_1$ and $r_2$ when $m_F=1$. The different $F$-parts of $r_1$ and $r_2$ have been colour-coded to visualise their contributions to each hybridized dressed state. The diagonalization for the $m_F$=1 ladder yields five eigenenergies with separations and multiplicities as annotated.}
    \label{fig:jbas}
\end{figure}
\subsubsection{Autler-Townes splittings for a type-I system}
For an ideal four-level ladder system, the AT-splitting that emerges in the EIT spectrum is directly proportional to the amplitude of the RF-field dressing the topmost two Rydberg states and the constant of proportionality is given by a combination of fundamental constants and the transition dipole moment between the Rydberg states.
Because of the quasi-one-electron character of Rydberg atoms, the latter can be calculated to very high accuracy. Early work, including Ref. \citenum{Holloway2014}, however, has introduced the misconception that for a type-\ref{eq:typeI} ladder with linear co-polarized fields, one is dealing with a single transition dipole moment between $r_1$ and $r_2$, and this has been reiterated pervasively in the literature~\cite{Fan2015a,Simons2016,Simons2016a,Holloway2017a,Zhang2019,Monika2020,Jing2020,Meyer2021a,Zhao2023,Bussey2024,Yang2024,Wu2025,Feng2025}. In fact, two transition dipole moments are in play differing by a factor of $\sqrt{2/3}\approx 0.8$ [see \fref{fig:r1r2}(a)] and, as noted in our above analysis, one will \textit{not} be measuring a pure AT-doublet but a four-peaked spectrum. 
\subsubsection{Type-I system: manifestation of a non-doublet AT-splitting\label{sec:nondoublet}}
 \Fref{fig:fourfive}(a) presents the development in AT splitting when moving our RF emitter up close ($\sim 2$~cm) to the vapor cell (see appendix \ref{app2}). To elucidate that each sidelobe in fact consists of two underlying peaks, we focus our attention on the blue-detuned branch ($\Delta_c>0$), where we note how the split-out peak becomes lopsided for high RF powers [see \fref{fig:fourfive}(b)]. If the goal is to establish an accurate electrometry standard based on a type-I ladder system, the four-peak nature should be accounted for because the splitting of each doublet-pair tunes at different rates: the inferred field amplitude will be skewed if an unresolved four-peak spectrum is interpreted as a pure doublet.
\Fref{fig:fourfive}(c) shows a simulated spectrogram for a type-\ref{eq:typeI} system with Doppler broadening ``turned off'', from which \fref{fig:fourfive}(d) picks out the co-polarized-fields spectrum that underpins the skewing of the experimental high power trace in \fref{fig:fourfive}(b). 

\subsubsection{Apparent angle-dependent peak positions}
The spectrogram in \fref{fig:fourfive}(c) underscores the important point that for linearly-polarized fields, the eigenenergy corresponding to a split-out peak does not change with $\theta$: the spectral features are straight vertical lines. While the position of a peak remains fixed, its prominence in the spectrum will, however, generally change. For two overlapping broadened peaks, this may give the appearance of a single feature that moves continuously as $\theta$ changes. This effect can be observed as wobbling of the side-lobes in the type-\ref{eq:typeI} column of \fref{fig:simvsexp}. 
\subsection{Allowed type-I transitions and the case of co-polarized fields	}
To be clear, it is incorrect to talk about ``the'' dipole moment between $D_{5/2}$ and $P_{3/2}$ levels for a linearly-polarized dressing field \cite{Rawat2020,Meyer2021a,Zhao2023,Feng2025,Jiang2025, Xie2025,Giat2025,Manchaiah2026}, because it depends on whether $m_J=|1/2|$ or $m_J=|3/2|$. It is also incorrect to resort to using the $m_J=|1/2|$ value \cite{Fan2015a,Yang2024,Simons2016a,Holloway2017a,Brown2023}, based on a rationale of all fields being $\pi$-polarized. This would be applicable in the absence of hyperfine structure of the ground state [see \fref{fig:jbas}(a) and (b)], in which case the system is described by two parallel, ideal four-state systems. For an atom in a specific, resolved hyperfine state we are, however, not dealing with parallel four-state systems, even when driving with co-polarized fields. This is evident from the diagrams in Figs.~\ref{fig:hyperfine-wide}(b) and \ref{fig:pitrans2}: moving up from any $m_F$-state of the ground level, one counts more than three transition-arrows. Moreover, a vertical cascade corresponding to a given $m_F$ [\fref{fig:jbas}(c) shows the example of $m_F=1$ from \fref{fig:hyperfine-wide}(b)] cannot be broken down into individual ``unique contributions'' [see \fref{fig:jbas}(d)] as was recently stated in Ref.~\citenum{Schlossberger2024a}.

The claim of Ref. \citenum{Schlossberger2024a} that the EIT scheme acts to probe the states of $r_1$ individually is invalidated by the fact that these states are degenerate (the hyperfine splitting of the Rydberg levels is negligible). They are therefore indirectly coupled with each other by the RF-field via the states of $r_2$ \cite{Chilcott2026}. Instead, when treating the example of the $m_F=1$-cascade in the hyperfine basis, the splittings of the system probed via EIT are found by diagonalizing the entire $7\times7$ matrix that describes the resonant RF interactions between the hyperfine states of $r_1$ and $r_2$. The result of this procedure is illustrated in \fref{fig:jbas}(e). As it should, it provides the same $m_F=1$ dressed energies encountered in the sequential procedure \fref{fig:r1r2}, where the eigenenergies are first found in the $J$-basis, after which the system is recast in the $F$-basis.

The insight that the co-polarized cases of Figs.~\ref{fig:hyperfine-wide}(b) and \ref{fig:pitrans2} are not represented by a collection of four-level systems means that there is \textit{a priori} no reason why they should give rise to simple AT doublets. Indeed, as we have shown, the type-\ref{eq:typeII} system features a prominent central peak in its spectrum. Interestingly, the two sidelobes of the type-\ref{eq:typeII} triplet tune as an ideal AT doublet with RF power with a tuning rate described by a single transition matrix element. The perfect absence of a central peak of the $\theta=0^\circ$ type-\ref{eq:typeI} system in \fref{fig:hyperfine-wide}(b) can be interpreted as a destructive quantum mechanical interference between the multiple parallel pathways represented by \eref{eq:3j}. Regardless, the spectrum does not represent a simple AT doublet. 
\subsection{Complementary spectroscopic response of type-I and type-II ladders}
Our recorded polarization spectrograms in \fref{fig:full-scans-transmission} for type-\ref{eq:typeI} and type-\ref{eq:typeII} ladders make the complementary response of these two systems compellingly clear. In particular, their transmissions recorded as a function of $\theta$ for $\Delta_\text{c}=0$ display out-of-phase oscillations reminiscent of the powers transmitted by and reflected off an optical polarization beam splitter. Unfortunately, because the type-\ref{eq:typeII} oscillation rides on a significant offset, the two ``output ports'' of our Rydberg atomic sensor do not serve to establish a general balanced RF polarimeter~\cite{Patterson2015} that would offer an absolute determination of $\theta$ from measurements at a single angle.
\section{Conclusion and outlook}
In summary, we investigated two types of transition ladders with complementary angular response. In particular, for all fields addressing the systems being co-linearly polarized, EIT spectroscopy of a type-\ref{eq:typeII} ladder displays a prominent spectral peak at $\Delta_c=0$, while the $\Delta_c=0$-signal for a type-\ref{eq:typeI} ladder is identically zero. This happens despite an apparent similarity and $\pi$-transitions being driven throughout the systems.

In the current study, we have focused on the case of linearly-polarized optical and RF fields, and we have shown that either of the two complementary spectroscopic signatures for type-\ref{eq:typeI} and -\ref{eq:typeII} ladders could serve to determine the angle of polarization for the RF field. The simultaneous high fidelity of our density matrix simulations for these complementary test cases instils confidence in our simulation framework's ability to synthesize the probe light transmission for more complex combinations of polarization states. In the future we plan to pursue the determination of arbitrary polarization states of the incoming RF field through machine learning approaches where the polarization states of probe and coupling light are varied to acquire a sequence of spectrograms. For this purpose the data diversity \cite{Gong2019} provided by the profoundly different response from type-\ref{eq:typeI} and type-\ref{eq:typeII} may provide a significant advantage. This would open up a new paradigm for quantum-metrological \cite{Nawrocki2019} polarimeters that at its heart is enabled by the quantisation of atomic angular momentum.

Implementations of Rydberg atomic electrometers found in the literature typically favour type-\ref{eq:typeI} ladders over type-\ref{eq:typeII} ladders and \fref{fig:setup}(d) reveals why this might be so. For a type-\ref{eq:typeII} ladder, the EIT-probed spectrum is dominated by a strong central peak, no matter the polarization angle. In contrast, a type-\ref{eq:typeI} ladder displays no central peak for co-polarized fields ($\theta=0$): the spectrum contains only split-out parts. An insight of this article is that these split-out parts do not form a simple AT doublet as encountered for an ideal four-level system [the $\theta=0^{\circ}$-case of \fref{fig:mickey-mouse}(d)]. Due to broadening mechanisms, the effect on experimentally measured spectra may or may not be subtle depending on the RF power in play. Also, depending on the application, the implications may or may not be important. For example, for Rydberg atomic radio receivers in communication \cite{Deb2018,Meyer2018,Song2019}, this is likely to have little practical consequence. On the other hand, for metrological field probes with SI-traceability as a selling point, accounting for the multivalued eigenenergy spectrum would seem inevitable if striving for the ultimate in accuracy.

\textit{Note added.}---After completing and submitting this work, we conducted a study where we used a horn antenna to dress the ${49D_{5/2}}\leftrightarrow{50P_{3/2}}$-transition with a resonant $\sim18$~GHz microwave field \cite{Chilcott2026a}. \Fref{fig:newexp} \cite{suppcite} presents the EIT spectrum for $\theta=90^\circ$, which clearly shows five distinct peaks, corroborating the non-doublet AT-splitting for a type-\ref{eq:typeI} ladder discussed in section \ref{sec:nondoublet}. 

\begin{acknowledgments}
	\noindent This work was supported by the Marsden Fund of New Zealand (Contracts No. UOO1923 and UOO1729) and by MBIE (Contract No. UOOX1915). We thank Jim Shaffer and Sebastian Hofferberth for their input on density matrix calculations at the beginning of this project. N.K. thanks Mikkel Andersen for discussions.
\end{acknowledgments}
M. Cloutman performed experiments with support from M. Chilcott, A. B. D., and S. O.; M. Cloutman analysed the data with support from M. Chilcott; M. Chilcott and A. E. carried out density matrix calculations; N. K., M. Cloutman, and M. Chilcott wrote the manuscript with input from all authors; N. K. conceived and supervised the project.

\appendix\section{EIT spectroscopy}\label{app1}
All calculations of transition frequencies were performed using ARC \cite{Sibalic2017}. The coupling laser was calibrated by scanning its frequency $\sim500$~MHz across the transitions from $P_{3/2}(F=3)$ to the $nD_{3/2}$ and $nD_{5/2}$ Rydberg states generating two EIT peaks; here $n=34$ or $n=35$ cf. \fref{fig:setup}(c). The probe and coupling beam powers were \SI{0.5}{mW} and \SI{300}{mW} respectively, and their $1/e^2$-diameters were \SI{2.0}{mm} and \SI{1.9}{mm}. To enable lock-in detection, the amplitude of the coupling beam was modulated at \SI{10}{\kilo\hertz} with an optical chopper. Probe and coupling beams were both linearly polarized in the horizontal plane after passing through polarizing beam splitter cubes and entered the vapor cell in a counter-propagating configuration to suppress Doppler broadening.
\vspace{4mm}

\section{RF dressing field}\label{app2}
The $\sim 55$~GHz RF field dressing our Rydberg atoms was produced by a photomixing antenna (Toptica Tera{-}Beam 1550). Two $\SI{\sim1550}{nm}$ fiber-coupled DFB lasers are combined to produce a beatnote with frequency $f_{\text{RF}}$ on an InGaAs photoconductive chip mounted on a bowtie-shaped antenna. The photomixing process produces linearly-polarized radiation at $f_{\text{RF}}$, which is collimated to a beam diverging at \SI{12}{\degree} by a silicon lens. The frequencies of the DFB lasers are temperature tuned, allowing us to target the $34\mathrm{D}_{5/2}\tranzs 35\mathrm{P}_{3/2}$ and $35\mathrm{D}_{3/2}\tranzs 36\mathrm{P}_{1/2}$ transitions at \SI{56.714}{GHz} and \SI{53.840}{GHz}, respectively. The emitter is positioned \SI{15}{cm} above the vapor cell, which is mounted on a wooden post to minimize metallic interference. Running the experiment with the RF dressing field tuned to a $S_{1/2}\leftrightarrow P_{1/2}$ transition, we verify that this gives rise to a spectrogram with the two straight horizontal lines \cite{Cloutman2024}. This ascertains that RF-radiation reflected back onto the cell is insignificant: reflected RF radiation field can change the resulting polarization state at the atoms and bend the lines in the $S_{1/2}\leftrightarrow P_{1/2}$-spectrogram \cite{Chopinaud2024}.

For the measurements presented in \Fref{fig:fourfive}(b) and (c), the emitter was positioned very close (\SI{<2}{\centi\metre}) to the cell. A \SI{0.75}{\milli\metre}-pitch wire grid polarizer was mounted between the emitter and the cell to suppress effects of a complex RF field polarization \cite{Chopinaud2024} and ensure a linear polarization at the position of the atoms.
\vspace{4mm}

\section{Density matrix calculations}\label{app3}
Polarization spectrograms (see \fref{fig:fourfive}) are simulated with QuTiP \cite{Johansson2013} by finding the steady-state atomic density matrix, $\rho_{\rm{ss}}$, under the combined action of the three fields; RF, coupling (c), and probe (p). 
The evolution of the density matrix, $\rho$, is governed by the Lindblad master equation \cite{Breuer2007},
\begin{equation}\label{eq:Lindblad}
    \dot{\rho} = \frac{1}{\ii\hbar}\left[H, \rho \right] + \sum_k \frac{\gamma_k}{2} \left(2 L_k \rho L_k^\dagger - \{L_k^\dagger L_k, \rho\}\right).
\end{equation}
The atomic Hamiltonian, $H$, in the dipole and rotating-wave approximations~\cite{Berman2011}, takes the block form,
\begin{equation} \label{eq:block}
    H = \frac{\hbar}{2}\begin{bmatrix}
        0 & \Omega_\text{p} & 0 & 0 \\
        \Omega_\text{p}^* & -2\delta_\text{p} & \Omega_\text{c} & 0 \\
        0 & \Omega_\text{c}^* & -2(\delta_\text{p} + \delta_\text{c}) & \Omega_\text{RF}\\
        0 & 0 & \Omega_\text{RF}^* & -2(\delta_\text{p} + \delta_\text{c} + \delta_\text{RF})
    \end{bmatrix}.
\end{equation}
Here the polarizations and Rabi frequencies of the probe, coupling and RF fields are encoded in the coupling operators $\Omega_k$ between manifolds, represented in (\ref{eq:block}) using an ordered basis of the $g$, $i$, $r_1$ and $r_2$ levels. The detuning blocks $\delta_k$ are scalar matrices because the individual states acted on by each block are not resolved. The problem is described in the $\ket{F m_F}$ basis, to deal with the fact that the probe laser field resolves transitions between hyperfine states. The second, dissipative term of \eref{eq:Lindblad} is composed of collapse operators $L_i$ with corresponding decay rates $\gamma_i$. A representation of $L_i$ for radiative decay in the $\ket{F m_F}$ basis is given in Supplementary Note 2 \cite{suppcite}. This term facilitates the inclusion of two classes of state decay: spontaneous emission, which has an operator for each possible polarization of the radiated photon, and incoherent decay, which is modelled with an auxiliary ``dummy state'' \cite{SedlacekThesis}. Incoherent decay includes transit-time broadening of all states, but it also describes some radiative decay of the Rydberg states, which usually involves the emission of multiple photons on a decay path outside the state space of the simulation. The ``dummy state'' then decays rapidly, with equal probability, to each sublevel of the ground-level manifold.

Unlike Ref.~\citenum{SedlacekThesis}, which time-evolved the system's Master equation from an equally-populated ground-state manifold, we directly solve for the steady state of the system. This entails finding the non-trivial zero of the Liouvillian superoperator $\mathcal{L}$ by recasting the density matrix as a vector (``vec-ing'') \cite{amshallem2015,Schlimgen2022}, and solving the equation $\dot{\rho}_{\rm{ss}} = \mathcal{L}\rho_{\rm{ss}} = 0$, using standard linear algebra techniques.

While the counter-propagating arrangement of coupling and probe beams compensates for the Doppler effect, we include the residual Doppler broadening of the room-temperature vapor by performing simulations with detunings that correspond to a range of atomic velocities along the probe/coupling beams, and averaging them according to the Maxwell-Boltzmann distribution. Our lock-in detection scheme is accounted for by removing a constant offset in the simulation: the resonant extinction of the probe beam without the coupling field.

\bibliography{EHF,references}
%\bibliographystyle{trybib}

%  at
% \href{http://www.physics.otago.ac.nz/staff_files/nk/files/krbsupp.pdf}{www.physics.otago.ac.nz/staff\_files/nk/files/krbsupp.pdf}
% %\href{http://www.physics.otago.ac.nz/staff_files/nk/files/MovieS1.mp4}{www.physics.otago.ac.nz/staff_files/nk/files/}
%\end{document}
\newpage
\clearpage
\setcounter{equation}{0}
\setcounter{figure}{0}
\setcounter{table}{0}

\renewcommand{\theequation}{S\arabic{equation}}
\makeatletter
\normalsize
%\includepdf[pages={1,{},{},2,{},3,{},4},pagecommand={},width=0.95\textwidth]{supparxiv.pdf}
\onecolumngrid
\setcounter{page}{1}
\setcounter{figure}{0}
\setcounter{table}{0}
\setcounter{page}{1 }
\renewcommand{\theequation}{S\arabic{equation}}
\renewcommand{\thefigure}{S\arabic{figure}}
% \renewcommand{\figurename}{\textbf{Supplementary Fig.}}
%\includepdf[pages={{1},-}]{collidersupp.pdf}

%{\scriptsize \noindent Supplementary Information for R. Thomas, M. Chilcott, E. Tiesinga. A. B. Deb, and N. Kj{\ae}rgaard: ``Observation of bound state self-interaction in a nano-eV atom collider'', \href{https://arxiv.org/pdf/1807.01174v2}{arXiv:1807.01174v2} (2018)}
\begin{center}{{\bf SUPPLEMENTAL MATERIAL}}\end{center}
\section*{Supplementary Note 1: Coupling from a state of the intermediate level to a dressed Rydberg state}
%\subsection*{}
To arrive at \eref{eq:3j} we must evaluate $\left|\sum_{F=1}^4 C^{Fm_F}_{Jm_JIm_I} \braket{F'm_F'}{r_0}{Fm_F}\right|^2$.

Using the Wigner Eckart theorem
we have 
\begin{eqnarray}
	\sum_{F=1}^4 C^{Fm_F}_{Jm_JIm_I} \braket{F'm_F'}{r_0}{Fm_F}=\sum_{F=1}^4 C^{Fm_F}_{Jm_JIm_I}C^{F'm_F'}_{Fm_F10} (2F'+1)^{-\frac{1}{2}}\ReducedMat{F'}{r_0}{F}\\\overset{\textrm{\rm{Ref.{\footnotesize\citenum{Sobelman1992}} p.84}}}{=}\sum_{F=1}^4 C^{Fm_F}_{Jm_JIm_I}C^{F'm_F'}_{Fm_F10} (2F'+1)^{-\frac{1}{2}}(-1)^{J'+I+F+1}\sqrt{(2F'+1)(2F+1)}\wsixj{J'}{F'}{I}{F}{J}{1}\ReducedMat{J'}{r_0}{J}\\=(-1)^{J'+I+F'+1}\sqrt{2F'+1}\ReducedMat{J'}{r_0}{J}\sum_{F=1}^4 	C^{Fm_F}_{Jm_JIm_I}C^{Fm_F}_{F'm_F'10} \wsixj{J'}{F'}{I}{F}{J}{1}\\\overset{\textrm{Ref. \citenum{Varshalovich1988} p.261}}{=}(-1)^{J'+I+F'+1}\sqrt{2F'+1}\ReducedMat{J'}{r_0}{J} 	C^{Im_I}_{J'-m_JF'm_F'}C^{10}_{J'-m_JJm_J},
\end{eqnarray}
	from which it follows that
	\begin{equation}
		\left|\sum_{F=1}^4 C^{Fm_F}_{Jm_JIm_I} \braket{F'm_F'}{r_0}{Fm_F}\right|^2\propto \begin{pmatrix}
    J' & J & 1\\
    -m_J & m_J & 0
    \end{pmatrix}^2\begin{pmatrix}
    J' & F' & I\\
    -m_J & m_F' & m_J-m_F'
    \end{pmatrix}^2.
	\end{equation}

\section*{Supplementary Note 2: Density matrix simulations}
In this supplement, we provide additional details on computing the density matrix, and connecting it to experimental observations. In particular, we include the matrix form of the operators necessary to formulate the master equation \eref{eq:Lindblad}, expanded over the hyperfine $\ket{F m_F}$ basis. The Python computer code used to produce the simulated results of \fref{fig:simvsexp} is available as part of this submission.

\subsection*{Field-induced Couplings}
The coupling operator $\Omega_k$ for each field ($k = p,c,\text{RF}$) is given by,
\begin{equation}
    \Omega_k = -\frac{\vec{\mu}_k \cdot \vec{E}_k}{\hbar},
\end{equation}
where $\vec{E}_k$ is the electric field, and $\vec{\mu}_k$ is a vector operator describing the dipole moments between states driven resonantly.

To populate the coupling matrix \eref{eq:block}, we make use of a spherical basis expansion~\cite{Auzins2014}. The spherical basis components $\hat{E}^{(1)}$ of the unit vector along the electric field direction (i.e., $\vec{E} = |\vec{E}| \hat{E}$), is given in terms of the Cartesian components by
\begin{align}
    \hat{E}^{(1)} &= -\frac{1}{\sqrt{2}}(\hat{E}_x + \ii \hat{E}_y),\\
    \hat{E}^{(0)} &= \hat{E}_z,\\
    \hat{E}^{(-1)} &= \frac{1}{\sqrt{2}}(\hat{E}_x - \ii \hat{E}_y).
\end{align}
Physically, $q = -1, 0, +1$ represents the change in angular momentum associated with $\sigma^-, \pi,$ and $\sigma^+$ excitations, respectively. 
Similarly, we expand $\vec{\mu}_k$ over the spherical basis $\hat{\epsilon}_{q}$, to introduce the angular component matrices $u_k^{(q)}$,
\begin{equation}
    %\vec{\mu}_k =e \left<n'L'||r||nL\right> \sum_{q = -1,0,1} \hat{\epsilon}_{q} u_k^{(q)}.
    \vec{\mu}_k \propto \sum_{q = -1,0,1} \hat{\epsilon}_{q} u_k^{(q)}.
\end{equation}
Using the spherical basis expansions, we factor the coupling operators into a scalar radial component $\Omega^{(r)}_k$ (the radial Rabi frequency) and an angular component,
\begin{equation}
    \Omega_k = \Omega_{k}^{(r)} \sum_{q=-1,0,1} \hat{E}^{(q)}_{k} u_{k}^{(q)}, \label{eqn:coupling_dfn}
\end{equation}
where the radial Rabi frequency between levels of principle quantum numbers $n$ and $n'$, and orbital angular momenta $L$ and $L'$ is 
\begin{align}
%    \Omega^{(r)}_k &= |\vec{E}| e \left<n'L'||r||nL\right>,\\
    \Omega^{(r)}_k & = |\vec{E}| e \int_{0}^{\infty} r^2\text{d}r \; \psi_{n'L'}^*(r) r \psi_{nL}(r),
\end{align}
in terms of the radial wavefunctions $\psi(r)$. We calculate this integral using the ARC~\cite{Sibalic2017}.

In the hyperfine basis, the elements of each matrix $u_{k}^{(q)}$ in \eref{eqn:coupling_dfn} are,
\begin{equation}\label{eq:angular_matrix}
    \begin{split}
    &\bra{S L J F m_F} u_{k}^{(q)} \ket{S' L' J' F' m_F'} =\\
    &\qquad(-1)^{I+S+J+J'+F+F'-m_F'}\\
        &\qquad\times \sqrt{(2L+1)(2L'+1)(2J+1)(2J'+1)(2F+1)(2F'+1)}\\
        &\qquad\times \begin{pmatrix}L&1&L'\\0&0&0\end{pmatrix}\begin{Bmatrix}L'&J'&S\\J&L&1\end{Bmatrix} \\
        &\qquad\times \begin{Bmatrix}J'&F'&I\\F&J&1\end{Bmatrix}\begin{pmatrix}F'&1&F\\-m_F'&q&m_F\end{pmatrix}.
    \end{split}
\end{equation}

\subsection*{Decay operators}
Decays that maintain coherence between states correspond to the emission of a photon via  $\sigma^+$, $\pi$, and $\sigma^-$ transitions, associated with atomic angular momentum changes $q = -1, 0, 1$, respectively. A decay path between two states is described by a collapse operator, $L_{k}^{(q)}$, for each $q$, with elements
\begin{equation}\label{eq:jump_op}
    \begin{split}
    &\bra{S L J F m_F} L_{k}^{(q)} \ket{S' L' J' F' m_F'} ^ 2= \\
    &\qquad (2L+1)(2J+1)(2J'+1)(2F+1)(2F'+1)\\
&\qquad\times \begin{Bmatrix}L'&J'&S\\J&L&1\end{Bmatrix}^2\times \begin{Bmatrix}J'&F'&I\\F&J&1\end{Bmatrix}^2\\
        &\qquad\times \begin{pmatrix}F' & 1 & F\\-m_F' & q &m_F\end{pmatrix}^2.
    \end{split}
\end{equation}
The right-hand side of this equation is proportional to the squared right-hand side of \eref{eq:angular_matrix}, with the notable feature that, for a given initial state, the sum over all $q$ and possible jumps is unity, so the decay path's $\gamma_k$ is the total radiative rate.

In our system, the decay from the Rydberg states is dominated by the non-radiative transit-time broadening~\cite{PhysRevA.22.2115} with a rate $\gamma_\text{TTB} \approx 2\pi \times \SI{200}{\kilo\hertz}$. Additionally, the vast majority of possible radiative Rydberg decay paths pass through states not included in our model. These (effectively) incoherent decay paths are modelled by means of a ``dummy state'' with no angular momentum character. Excited states decay to this dummy state via a collapse operator with a single non-zero element. In the case of the $i$-level states, the rate is given by the transit-time broadening. For the two Rydberg levels, the total radiative decay rate to states other than $\left|i\right>$ is added to this. The ``dummy state'' itself decays rapidly into the $\ket{g}$ manifold (with a near-infinite $\gamma$). We note the rudimentary model for transit-time broadening used is valid for large beams where the radiative decay rate is larger than the transit rate, and this system's dominant radiative decay process is the $\gamma_i \approx 2\pi \times \SI{6}{\mega\hertz}$.

\subsection*{Probe extinction}

As discussed in the main text, the extinction of the probe beam of polarization $q$ through an atomic vapor follows the Beer-Lambert law, \eref{eq:beer}. The extinction coefficient for a $q$-polarized field, $\alpha^{(q)}$, through a vapor of density $n$ at a transition of frequency $\omega_p$ is composed of off-diagonal elements of the atomic density matrix $\rho$,
\begin{equation}\label{eq:alpha}
  \alpha^{(q)} = \frac{2n\omega_p }{c \epsilon_0 \hbar } \sum_{g_l \in g, i_m \in i} \left(\mu^{(q)}_{i_m g_l}\right)^2\text{Im}\left(\frac{\rho_{g_l i_m}}{\Omega_{g_l i_m}}\right),
\end{equation}
with the sum running over the $q$-coupled states of the $g$- and $i$-level manifolds. The density matrix is the steady state of the  master equation of \eref{eq:Lindblad} with the atomic Hamiltonian \eref{eq:block}, and the operators described above.

\section*{Supplementary Figure}
\begin{figure}[h!]
    \centering
\includegraphics[width=0.6\columnwidth]{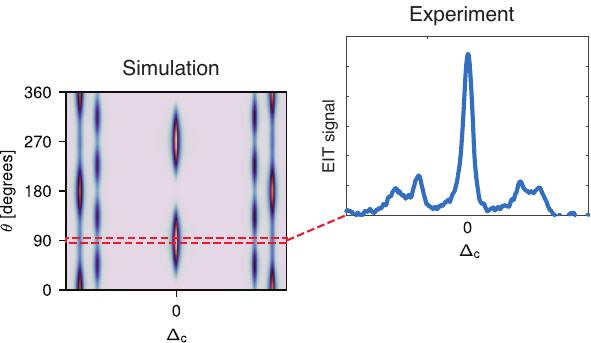}
    \caption{EIT spectrum recorded for a linearly polarized microwave field ($\theta=90^\circ$) dressing the ${49D_{5/2}}\leftrightarrow{50P_{3/2}}$-transition with a resonant $\sim18$~GHZ microwave field \cite{Chilcott2026a}. The experimental  data shows a clear splitting of each side-lobe.}
    \label{fig:newexp}
\end{figure}
\renewcommand{\refname}{\textbf{Supplementary References}}

\end{document}